\documentclass[epjST]{svjour}
\usepackage{graphicx}
\usepackage[
backend=biber,
hyperref=true,
sorting=none,
language=english,
style=phys
]{biblatex}
\usepackage[english]{babel}
\usepackage[utf8]{inputenc}
\usepackage{microtype}          
\usepackage[breaklinks=true,colorlinks=true,linkcolor=blue,urlcolor=blue,citecolor=blue]{hyperref}

\usepackage{xcolor}     %
\usepackage{mathtools}
\mathtoolsset{showonlyrefs=false}
\usepackage{braket}
\usepackage{amssymb}
\usepackage{bm}%
\usepackage{enumerate}

\DeclareFieldFormat[article]{title}{}

\newcommand{\refeqn}[1]{Eq.~\eqref{#1}}

\newcommand{\eqrefs}[2]{Eqs.~\eqref{#1} and \eqref{#2}}

\newcommand{\lowerbossphantom}{\vphantom{\bar{\bar{x}}}}
\newcommand{\upperbossphantom}{\vphantom{\dagger}}
\newcommand{\tempop}[3][\textstyle]{\settowidth{\dimen1}{$#1\hat{#2}$}\makebox[\dimen1][l]{$#1\hat{#2\mspace{#3}}$}}
\newcommand{\xop}[1]{{\mathchoice{\tempop[\displaystyle]{#1}{3.5mu}}{\tempop{#1}{3.5mu}}{\tempop[\scriptstyle]{#1}{3.5mu}}{\tempop[\scriptscriptstyle]{#1}{3mu}}}}
\newcommand{\chat}[1]{\ensuremath{\xop{#1}}}
\newcommand{\aop}[2]{\ensuremath{\chat{c}_{#1#2\lowerbossphantom}^{\upperbossphantom}}}
\newcommand{\cop}[2]{\ensuremath{\chat{c}_{#1#2\lowerbossphantom}^{\dagger\upperbossphantom}}}

\renewcommand{\i}{\mathrm{i}}
\renewcommand{\d}{\mathrm{d}}
\newcommand{\tn}[1]{\textnormal{#1}}

\begin{document}
\title{L\"owdin's symmetry dilemma within Green functions theory for the one-dimensional Hubbard model}
\author{J.-P. Joost\inst{1} \and N. Schlünzen\inst{1}  \and S. Hese\inst{1} \and M. Bonitz\inst{1}\fnmsep\thanks{\email{bonitz@physik.uni-kiel.de}}\and C. Verdozzi\inst{2}\and  P. Schmitteckert\inst{3} \and M. Hopjan\inst{4}}
\institute{Institute of Theoretical Physics and Astrophysics, Kiel University, Germany \and Physics Department and ETSF, Lund University, Sweden \and HQS Quantum Simulations GmbH, Karlsruhe, Germany \and Institute for Theoretical Physics, Göttingen University, Germany}
\abstract{
The energy gap of correlated Hubbard clusters is well studied for one-dimensional systems using analytical methods and density-matrix-renormalization-group (DMRG) simulations. Beyond 1D, however, exact results are  available only for small systems by quantum Monte Carlo. For this reason and, due to the problems of DMRG in simulating 2D and 3D systems, alternative methods such as Green functions combined with many-body approximations (GFMBA), that do not have this restriction, are highly important. However, it has remained open whether the approximate character of GFMBA simulations prevents the computation of the Hubbard gap. Here we present new GFMBA results that demonstrate that GFMBA simulations are capable of producing reliable data for the gap which agrees well with the DMRG benchmarks in 1D. An interesting observation is that the accuracy of the gap can be significantly increased when the simulations give up certain symmetry restriction of the exact system, such as spin symmetry and spatial homogeneity. This is seen as manifestation and generalization of the ``symmetry dilemma'' introduced by Löwdin for Hartree--Fock wave function calculations. %
} %
\maketitle
\section{Introduction}
\label{s:intro}

Symmetry and its possible violation or breaking are basic notions in our understanding of physical
phenomena. In essence, this is because the symmetry transformations in a physical system 
relate to conservation laws of specific observables.
 
Around six decades ago, L\"owdin introduced the term {\it symmetry dilemma}\footnote{Quoting Löwdin from that reference, p. 498: ``In my opinion, the Hartree--Fock scheme based on a single Slater determinant $D$ is in a dilemma
with respect to the symmetry properties and the normal constants of motion $\Lambda$. The assumption
that $D$ should be symmetry-adapted or an eigenfunction to h. leads to an energy $\langle H\rangle$ high above
the absolute minimum,...''} \cite{POL_RevModPhys} to portray a situation where imposing symmetry constraints in a Hartree--Fock (HF) calculation of the eigenfunction of a given electronic system gives an energy eigenvalue higher than in absence of such constraints. 
In other words, the removal of constraints increases the variational flexibility of the HF wavefunction, giving an approximate energy value closer to the exact one, but typically does not preserve the system's good quantum numbers and symmetry properties. A well-known example are HF simulations of the ground state of the uniform electron gas, e.g., Refs.~\cite{overhauser_PhysRevLett.4.462,bernu_cpp17,dornheim_physrep_18}.
From L\"owdin's original insight, extensive research has spurred, 
to develop approaches where symmetries in a single-determinant wavefunction are deliberately broken, and subsequently reintroduced via symmetry projection operators, to attain a variationally improved multi-determinant state (see, e.g., Ref.~\cite{Scuseria} for a recent discussion).

Ambivalence in the use of symmetry is in fact of very general occurrence, and concerns
both finite and infinite systems.  An interesting example is provided by spontaneous symmetry breaking (SSB) \cite{Anderson}. In a rigorous sense, SSB only takes place in the thermodynamic limit %
However, exact numerical evidence from finite systems shows that the ``disjointness'' (typical of macroscopic sizes) of phases or different values/orientations of the order parameter is replaced by a crossover behavior across finite barriers in Hilbert space,  whose sharpness and strength increases on enlarging the system's size. An example is Wigner crystallization in finite electron systems, such as quantum dots, e.g., Refs.~\cite{filinov_prl_1, filinov_pss_00,boening_prl_8}, for which also SSB in HF calculations was investigated \cite{landman_PhysRevLett.82.5325}; other examples are found in ultracold bosons in traps, nuclear matter, and quantum chemistry; for an overview, see Ref.~\cite{Yannouleas_2007}.
It can thus be methodologically expedient  
to artificially break the symmetry in a finite system, to
gain insight about the system behavior in the thermodynamic limit. An often used prescription
is the addition of small external sources lowering the symmetry \cite{Kirkwood,Bogolubov},
but under the stipulation that it is understood that true SSB occurs only asymptotically.

Another central element to consider in addressing the symmetry-related behavior of 
a system is electronic or inter-particle correlations. These have deep influence in various situations, e.g.
condensed-matter systems and materials, plasmas, nuclear matter, and cold atoms,
to mention a few %
\cite{Gebhard,bonitz_rpp_10,balzer_prb_9,stefanucci_cambridge_2013}.
Clearly, the interplay of 
electronic interactions and symmetry constraints affects the system's 
properties in a way that is not accountable for within a free-particle or mean-field picture.
It should be noted, though, that already within a wavefunction framework,  some theories 
going beyond the mean-field picture can 
mimic the effect of strong electronic correlations with wavefunctions that do not 
respect the expected symmetry (see, e.g., Ref.~\cite{Mendl}).

In this work, we take a different route from wavefunctions, and we study the effect of lifting/breaking symmetry in the presence
of significant electronic correlations within many-body perturbative Green functions theory. As the system of choice
with strong correlations, we 
consider the Hubbard model \cite{Hubbard1,Hubbard2,Gebhard} which,
 via a minimum-complexity Hamiltonian, describes
the key trends in the behavior of interacting electrons in the energy bands of a solid.
For this reason, Hubbard-like models have been applied in many contexts and to a wide typology of systems, both in and out of equilibrium, see, e.g., Refs.~\cite{HubbardModel,TARRUELL2018365,EcksteinPRL09,schluenzen_prb16,balzer_prb16,balzer_prl_18,joost_pss_18,bostrom_charge_2018,HopjanPRB18,BostromPSSB19,joost_19_nanolett,covito_real_2018,cohen_green_2020,kalvova_generalized_2018,kalvova_beyond_2019}.

Even though the Hubbard Hamiltonian is considerably simpler than that of a realistic material,
exact solutions for the Hubbard model are only known in special cases: in one dimension, an exact analytical 
treatment is possible via Bethe-ansatz techniques \cite{LiebWu}, and exact numerical solutions for finite samples
can be obtained via the density-matrix-renormalization-group (DMRG) method \cite{daley_04,vidal_prl_04,white_prl_04}
(including spin-charge separation effects~\cite{PS3,PS4,PS5}).
Using configuration interaction (CI) \cite{CI} or quantum Monte Carlo methods \cite{gull_continous_2011,Hirsch,MuramatsuEPL}, exact solutions can be obtained for any dimensionality, but only for small clusters. Finally, an exact description is 
possible in the limit of infinite dimensionality via dynamical-mean-field theory (DMFT) \cite{DMFT1,DMFT2,DMFT3}. 
In all other cases (notably, in two and three dimensions), some level of approximation must be
introduced. Yet, high accuracy can be attained, for example via the diagrammatic Monte Carlo technique 
\cite{Troyer}, or by extensions of DMFT via cluster \cite{DMFTclu} or diagrammatic approaches \cite{DMFTdia1,DMFTdia2}.

A premier method traditionally applied to the Hubbard model is the Green functions formalism combined 
with many-body perturbation theory (GFMBA)~\cite{Fetter,AGD}. The GFMBA method is a general framework
that can be used in any dimension (i.e. also for 2D and 3D Hubbard models), scales not too unfavourably with system's size, can deal
with both static and time-dependent regimes \cite{stefanucci_cambridge_2013,balzer_2013_nonequilibrium}, and is also practically viable for implementation
for realistic systems.  Furthermore, a very recent reformulation of the method in terms
of coupled one- and two-particle propagators \cite{schluenzen_19_prl,joost_20_prb}
has considerably increased its scope and range of applicability. 
In the case of the GFMBA method, correlation effects are included
via selected classes (possibly infinite sums) of diagrams in the selfenergy, or via truncated
iterative functional-derivative schemes. This leads to different perturbative treatments~\cite{AGD,Fetter}, 
e.g. HF, second-order Born approximation (SOA), third-order approximation (TOA) \cite{von_friesen_successes_2009,schluenzen_jpcm_19}, 
$GW$, and the particle--particle and particle--hole $T$-matrix approximation \cite{stefanucci_cambridge_2013,schluenzen_jpcm_19}. 
Comparisons against exact benchmarks for finite systems have shown that the GFMBA method works
well for not too strong interactions \cite{von_friesen_successes_2009,schluenzen_jpcm_19,schluenzen_prb17_comment,schluenzen_prb17}.

The GFMBA approach has recently been under extensive
scrutiny in relation
to the existence of multiple solutions in the ground state and a potential convergence to nonphysical ones
\cite{Lani,tandetzky_multiplicity_2015,gunnarsson_breakdown_2017} (for an example
from an out-of-equilibrium systems, see e.g. Ref.~\cite{Spataru}).
However, in the discussion the multiplicity of GFMBA solutions,
little or no attention has been given so far in the literature to the explicit role of symmetry, particularly
in relation to the Hubbard model. For the exact solution of this model,
important ground-state properties and symmetries are well known \cite{lieb_two_1989,lieb_uniform_1993} and
in common practice, these symmetries are granted by default also to
the approximate solutions. 

There is scarce knowledge on how these symmetries affect approximate 
treatments beyond mean-field theory.  
For example, for Hartree--Fock treatments of the Hubbard model, it is well known that a phase transition from a paramagnetic ground state to an antiferromagnetic one of unphysical nature occurs at a critical interaction, $U_\tn{c}$, where the specific value depends on the system size and geometry \cite{Hirsch,Penn}. Yet, at the same time, the (unphysical) broken spin-symmetry solutions result
in a ground-state energy closer to the exact one, as well as in the emergence of a band gap (the 
latter is absent within spin-symmetric/spin-restricted mean-field schemes). This raises the question whether selfenergy approximations beyond Hartree--Fock can be found that violate the symmetry properties of the exact solution as well,
and yet provide ``improved'' values of relevant observables, such as the ground-state energy or the Hubbard gap. In other words,

{\it 1) Is there a L\"owdin symmetry dilemma for the Hubbard model within many-body perturbation theory?\\
\indent 2) And in case, how is this related to solution multiplicity?}

Our answer to the first question is in the affirmative: By considering the Hubbard model 
in the one-dimensional case, and comparing GFMBA, CI, and DMRG results,
we find that, lifting the symmetry constraints artificially, simulates the effect of having 
more correlation effects in the system, and leads to a significant improvement of 
observables like the system's ground-state energy or the spectral functions, even at fairly large interactions.
In particular, our discussion will be especially focussed on the charge energy gap $\Delta$ (also known as Mott-Hubbard gap). For a system containing $N$ particles it depends on the ground state energies of the $N$, $N+1$ and $N-1$ system,
\begin{align}\label{eq:gap}
    \Delta = E_\tn{GS}\left(N+1\right) + E_\tn{GS}\left(N-1\right) - 2 E_\tn{GS}(N)\,.
\end{align}
This is a central  quantity of the Hubbard model, related to the metal--insulator transition (MIT) at a characteristic interaction strength.\footnote{The characteristics of the MIT (and thus of the gap) depend on dimensionality (see, e.g., Refs.~\cite{BrinkmanRice,Gebhard}). For $D=1$,
the exact solution shows that the insulating phase (and the charge gap) exist for any $U>0$  (i.e. no MIT  occurs)
 \cite{LiebWu68}. For infinite dimensions, where the model is exactly solvable via DMFT, 
 the MIT occurs at finite interaction values \cite{DMFT3} (however, there is coexistence of metallic
 and insulating solutions in an interval $U_{\tn{c}_2}<U<U_{\tn{c}_1}$). This picture qualitatively remains
 in three dimensions, with however quantitative modifications, due to antiferromagnetic fluctuations \cite{MuramatsuEPL}.
 However, as proposed recently \cite{Toschi} %
 non-local (in particular, longe-range) antiferromagnetic correlations appear to play an even  more dramatic role in two dimensions, where the system is gapped, for any $U>0$, with absence of a MIT.
}

For the second question, we find that the occurrence of multiple (metastable) solutions 
is central to the connection between symmetry lifting and improved values
of certain observables in the Hubbard model.
Besides looking at what happens when lifting symmetry and why, we will also consider the 
possible implications in physical terms. However, due to the explorative nature of this initial study, 
we will not explore/discuss strategies to restore symmetry. This is left to future work. 

The paper is organized as follows: in Sec.~\ref{s:gf_theory}, we introduce the Hubbard Hamiltonian and the Green functions formalism for HF and SOA treatments within the three self-consistency protocols with different levels of symmetry constraints.  Sec.~\ref{s:hf} presents results from a mean-field treatment, specifically for the ground-state energy, the Hubbard gap, and the magnetic  moment as function of the interaction strength $U$.
In Sec.~\ref{s:soa} we consider the SOA, discussing at the same time
the ground-state density matrix and equilibrium spectral function for all the three  self-consistency prescriptions.
The multiple solutions in SOA are further analyzed in  Sec.~\ref{s:multi} in terms of ground-state energy values 
and self-consistency convergence errors. 
Finally, Sec.\ref{s:gap} focuses on a characterization of the Hubbard gap, 
where HF and SOA results are compared to DMRG ones,
and the dependence on system size,  $L$, is taken into account.
Our conclusions and a brief outlook are presented in Sec.~\ref{s:conclusion}.

\section{Green Functions Theory}
\label{s:gf_theory}
We consider the Fermi-Hubbard model which is described by the Hamiltonian %
\begin{align}
 \chat{H} = -J\sum_{\left<i,j\right>} \sum_{\sigma=\uparrow,\downarrow} \cop{i}{,\sigma}\aop{j}{,\sigma} + U \sum_i \chat{n}_{i,\uparrow} \chat{n}_{i,\downarrow}\, ,
\label{eq:h-hubbard}
\end{align}
where $J$ is the hopping amplitude between adjacent lattice sites, and $U$ is the on-site interaction strength. The operators ${\chat{c}}_{i,\sigma}^\dagger$ and ${\chat{c}}_{j,\sigma}$ create and annihilate an electron with spin projection $\sigma$ at site $i$ and $j$, respectively, and the density operator is given by $\chat{n}_{i,\sigma} = {\chat{c}}_{i,\sigma}^\dagger {\chat{c}}_{i,\sigma}$.\\
The one-body nonequilibrium Green function of the system is defined by the canonical operators for complex times $z$ on the Keldysh contour $\mathcal{C}$~\cite{keldysh64,bonitz_pss_19_keldysh},
\begin{align}
    G_{ij,\sigma}(z,z')=\frac{1}{\i\hbar}\left\langle \mathcal{T}_\mathcal{C} \left\{\chat{c}_{i,\sigma}(z)\chat{c}^\dagger_{j,\sigma}(z')\right\} \right\rangle\,,
\end{align}
where $\mathcal{T}_\mathcal{C}$ is the time-ordering operator on the contour, and the averaging is performed with the correlated unperturbed density operator of the system. In the rest of this work we specialize to the equilibrium regime, i.e. when the system is not acted upon by external fields. In that case the real-time components~\cite{stefanucci_cambridge_2013,balzer_2013_nonequilibrium} of the Green function depend only on the difference of the two time arguments. Further, the retarded (R) component for the spin projection $\sigma$ obeys the Dyson equation
\begin{align}
 \bm{G}^{\mathrm{R}}_{\sigma}(\omega) = \bm{G}_0^{\mathrm{R}}(\omega) + \bm{G}_0^{\mathrm{R}}(\omega) \bm{\Sigma}^{\mathrm{R}}_{\sigma}(\omega) \bm{G}^{\mathrm{R}}_{\sigma}(\omega)\,,
 \label{eq:dyson}
\end{align}
where all quantities are matrices in the orthonormal single-particle basis $\{\ket{i}\}$. For spin-compensated situations, the noninteracting retarded Green function $\bm{G}_0^{\mathrm{R}}(\omega)$ is independent of the spin projection and given by
 \begin{align}
  G_{0,ij}^{\mathrm{R}}(\omega) = \bra{i}(\omega - \chat{h}_0 + \mathrm{i} \eta)^{-1}\ket{j}, \qquad \eta \to 0^+\,,\label{eq:G0}
 \end{align}
depending only on the single-particle contribution to the Hamiltonian
 \begin{align}
\chat{h}_0 = -J\sum_{\langle i,j\rangle} {\chat{c}}_{i,\uparrow}^\dagger{\chat{c}}_{j,\uparrow} = -J\sum_{\langle i,j\rangle} {\chat{c}}_{i,\downarrow}^\dagger{\chat{c}}_{j,\downarrow} \,.\label{eq:hamiltonian}
\end{align}
For numerical reasons a finite value of $\eta=10^{-2}$ is used throughout this work. If the exact selfenergy $\bm{\Sigma}^{\mathrm{R}}_\sigma(\omega)$ of the system was known, Eq.~(\ref{eq:dyson}) would provide the exact single-particle Green function. However, in practice many-body approximations to the selfenergy have to be used. 

In this work we employ the time-diagonal Hartree--Fock as well as the time-non-local second-order Born approximation. The retarded component of the HF selfenergy is defined as\footnote{Mind that in the Hubbard model all exchange contributions vanish and only the direct diagrams remain.}
\begin{align}
 \bm{\Sigma}^{\mathrm{R,HF}}_{\sigma}(t) = \delta(t,0)\, U\, \mathrm{diag}(n_{1,\bar\sigma},\ldots,n_{L,\bar\sigma})\,,\label{eq:sigmaHFR}
\end{align}
where $\delta$ is the Dirac delta function for the relative time $t$ in equilibrium and $\mathrm{diag}( \cdot )$ represents a diagonal matrix with the given arguments as diagonal entries.
Further, $\bar \sigma$ denotes the spin-projection opposite to $\sigma$, $L$ the number of lattice sites, and $n_{i,\sigma} = n_{ii,\sigma}$. The density matrix is given by the less component of the Green function,
 \begin{align}
      \bm{n}_\sigma = -\i\hbar\int^{\infty}_{-\infty}\frac{\d\omega}{2\pi}\bm{G}^{<}_{\sigma}(\omega)\,.
 \end{align}
For the correlated Green function the less ($<$) and greater ($>$) components can be determined by
 \begin{eqnarray}
  \bm{G}^{<}_{\sigma}(\omega) &=& -f_\mathrm{F}(\omega-\mu) \Big[\bm{G}^{\mathrm{R}}_{\sigma}(\omega) - \bm{G}^{\mathrm{A}}_{\sigma}(\omega) \Big],
  \label{eq:Gl}\\
  \bm{G}^{>}_{\sigma}(\omega) &=& \bar{f}_\mathrm{F}(\omega-\mu) \Big[\bm{G}^{\mathrm{R}}_{\sigma}(\omega) - \bm{G}^{\mathrm{A}}_{\sigma}(\omega) \Big],
  \label{eq:Gg}
 \end{eqnarray}
 with the Fermi function $f_\mathrm{F}(\omega) = 1 / \left(\tn{e}^{\beta\omega} + 1\right)$, $\;\bar{f}_\mathrm{F}(\omega) = 1-f_\mathrm{F}(\omega)$, the inverse temperature $\beta$, and $\bm{G}^{\mathrm{A}}_{\sigma}(\omega) = \left[\bm{G}^{\mathrm{R}}_{\sigma}(\omega)\right]^\dagger$.\\
For the SOA selfenergy the retarded component is given by
\begin{equation}
 \bm{\Sigma}^{\mathrm{R,SOA}}_\sigma(t) = \bm{\Sigma}^{\mathrm{R,HF}}_{\sigma}(t) +  \Theta(t)\Big[\bm{\Sigma}^{>,\mathrm{SOA}}_\sigma(t)-\bm{\Sigma}^{<,\mathrm{SOA}}_\sigma(t)\Big]\,,\label{eq:sigmar}
\end{equation}
with the Heaviside step function $\Theta(t)$ and the greater and less components of the selfenergy
\begin{equation}\label{eq:sigmagl}
 \bm{\Sigma}^{\gtrless,\mathrm{SOA}}_\sigma(t) = -(\i\hbar)^2 U^2(t) \bm{G}^{\gtrless}_{\sigma}(t) \circ \bm{G}^{\gtrless}_{\bar\sigma}(t) \circ \left[\bm{G}^{\lessgtr}_{\bar\sigma}(t)\right]^\dagger\,.
\end{equation}
Here, $\circ$ denotes the Hadamard product between matrices, and the $\bm{G}^\gtrless(t)$ are determined by the inverse Fourier transform,
\begin{align}
 \bm{G}^\gtrless_\sigma(t) = \mathcal{F}^{-1}\left[\bm{G}^\gtrless_\sigma(\omega)\right] \coloneqq \int_{-\infty}^{\infty} \frac{\mathrm{d}\omega}{2\pi} \,\mathrm{e}^{-\mathrm{i}\omega t} \bm{G}^\gtrless_\sigma(\omega) \,.\label{eq:FT}
\end{align}
Finally, using the Fourier transform, $\bm{\Sigma}^{\mathrm{R}}_\sigma(\omega) = \mathcal{F}\left[\bm{\Sigma}^{\mathrm{R}}_\sigma(t)\right]$, the selfenergies, \eqrefs{eq:sigmaHFR}{eq:sigmar}, can be included in Eq.~(\ref{eq:dyson}).\\
Since the selfenergy, in general, is a functional of the single-particle Green function, Eq.~(\ref{eq:dyson}) has to be solved iteratively for both spin components until a self-consistent solution is found.
The choice of a suitable initial value is crucial and can affect the final result of the iteration. Here, if not mentioned otherwise, $\bm{G}_0^{\mathrm{R}}(\omega)$ is chosen as the starting point. 

In summary, for the self-consistent solution of Eq.~(\ref{eq:dyson}) the following scheme is iterated until convergence is achieved: 
\begin{enumerate}[\indent1)]
\setcounter{enumi}{-1}
 \item Diagonalize the single-particle Hamiltonian $\chat{h}_0$, cf. Eq.~(\ref{eq:hamiltonian}), set $\bm{G}_0^{\mathrm{R}}(\omega)$ via Eq.~(\ref{eq:G0}) and choose an initial value to start the iteration, e.g. $\bm{G}^{\mathrm{R}}_\sigma(\omega) = \bm{G}_0^{\mathrm{R}}(\omega)$
 \item Calculate $\bm{G}^{\gtrless}_\sigma(\omega)$ from $\bm{G}^{\mathrm{R}}_\sigma(\omega)$, using Eqs.~(\ref{eq:Gl}) and (\ref{eq:Gg})
 \item Perform the inverse Fourier transform $\bm{G}^{\gtrless}_\sigma(t) = \mathcal{F}^{-1}\big[ \bm{G}^{\gtrless}_\sigma(\omega) \big]$, cf. Eq.~(\ref{eq:FT}) 
 \item Calculate $\bm{\Sigma}^{\gtrless}_\sigma(t)$ and $\bm{\Sigma}^{\mathrm{R}}_\sigma(t)$ using Eqs.~(\ref{eq:sigmaHFR}), (\ref{eq:sigmar}) and (\ref{eq:sigmagl})
 \item Perform the Fourier transform $\bm{\Sigma}^{\mathrm{R}}_\sigma(\omega) = \mathcal{F}\left[ \bm{\Sigma}^{\mathrm{R}}_\sigma(t) \right]$ 
 \item Solve the Dyson equation for $\bm{G}^{\mathrm{R}}_\sigma(\omega)$, Eq.~(\ref{eq:dyson}), using the new $\bm{\Sigma}^{\mathrm{R}}_\sigma(\omega)$
 \item If $\bm{G}^{\mathrm{R}}_\sigma(\omega)$ is not yet converged start again at 1)
\end{enumerate}
To improve the convergence of the above scheme, the input Green function at iteration $k$,  $\bm{G}^\mathrm{R}_{k,\mathrm{in}}$ (the spin index is neglected here), is determined by mixing the solutions of the two previous iterations
\begin{equation}
 \bm{G}^{\mathrm{R}}_{k,\mathrm{in}}(\omega) = \alpha \bm{G}^{\mathrm{R}}_{k-1,\mathrm{out}}(\omega) + (1-\alpha) \bm{G}^{\mathrm{R}}_{k-1,\mathrm{in}}(\omega)\,,\label{eq:mixing}
\end{equation}
where a mixing parameter of $\alpha=0.05$ is used. The error at iteration $k$ is given by
\begin{align}
    \epsilon_k = \frac{1}{\alpha L} \int^{\infty}_{-\infty}\frac{\d\omega}{2\pi} \Big|D_k(\omega) - D_{k-1}(\omega)\Big|\,,
\end{align}
where $D(\omega)$ is the density of states (DOS) of the system,
\begin{align}
    D(\omega) = \i\hbar \sum_{\sigma=\uparrow,\downarrow} \mathrm{tr} \Big[ \bm{G}^{>}_{\sigma}(\omega) - \bm{G}^{<}_{\sigma}(\omega) \Big]\,.
\end{align}
To ensure the convergence of the iteration scheme an error threshold of $\epsilon_\mathrm{thr}=10^{-12}$ is used in this work.
In addition to spectral properties, the single-particle Green function gives access to the total energy of the system~\cite{stefanucci_cambridge_2013,Fetter}%
 \begin{align}
      E_\mathrm{tot} = \frac{1}{2}E_\mathrm{kin} + E_\mathrm{GM}\,,
 \end{align}
which combines the kinetic part
 \begin{align}
      E_\mathrm{kin} = \sum_\sigma\mathrm{tr}\left(\bm{h}_0\bm{n}_\sigma\right)\,,
 \end{align}
 and the Galitskii--Migdal interaction energy
 \begin{align}
      E_\mathrm{GM} = -\frac{\i\hbar}{2}\sum_{\sigma=\uparrow,\downarrow}\int^{\infty}_{-\infty}\frac{\d\omega}{2\pi}(\omega-\mu)\,\mathrm{tr}\Big[\bm{G}^{<}_{\sigma}(\omega)\Big]\,,
 \end{align}
where, in the GFMBA simulations of the present paper, the chemical potential is set to $\mu=0$. %

In this paper, we consider three distinct cases for which different symmetry restrictions are imposed on the Green function during the solution of the Dyson equation:
\begin{enumerate}[(I)]
 \item ``uniform'' (uni): the system is required to be translationally invariant. In this case the iteration scheme is solved in momentum space where the Green function and selfenergy are diagonal, i.e. $G_{ij,\sigma}^{\mathrm{R}}(\omega) \rightarrow{} G_{p,\sigma}^{\mathrm{R}}(\omega)$.
 \item ``restricted spin''  (rs): the system is required to be spin-symmetric. In this case the iteration scheme is solved for one spin projection only since the Green function and selfenergy are spin-independent, i.e. $\bm{G}_{\uparrow}^{\mathrm{R}}(\omega) = \bm{G}_{\downarrow}^{\mathrm{R}}(\omega)$.
 \item ``unrestricted spin'' (us): no restrictions regarding both, the translation and spin symmetry are imposed. In this case an antiferromagnetic state is chosen to start the iteration.\footnote{When choosing the spin-symmetric $\bm{G}_0^{\mathrm{R}}(\omega)$, as the initial value, the iteration will not break spin symmetry.}
\end{enumerate}

\section{Spin Symmetry in the Mean-Field Approximation}
\label{s:hf}

\begin{figure}
 \includegraphics[width=\textwidth]{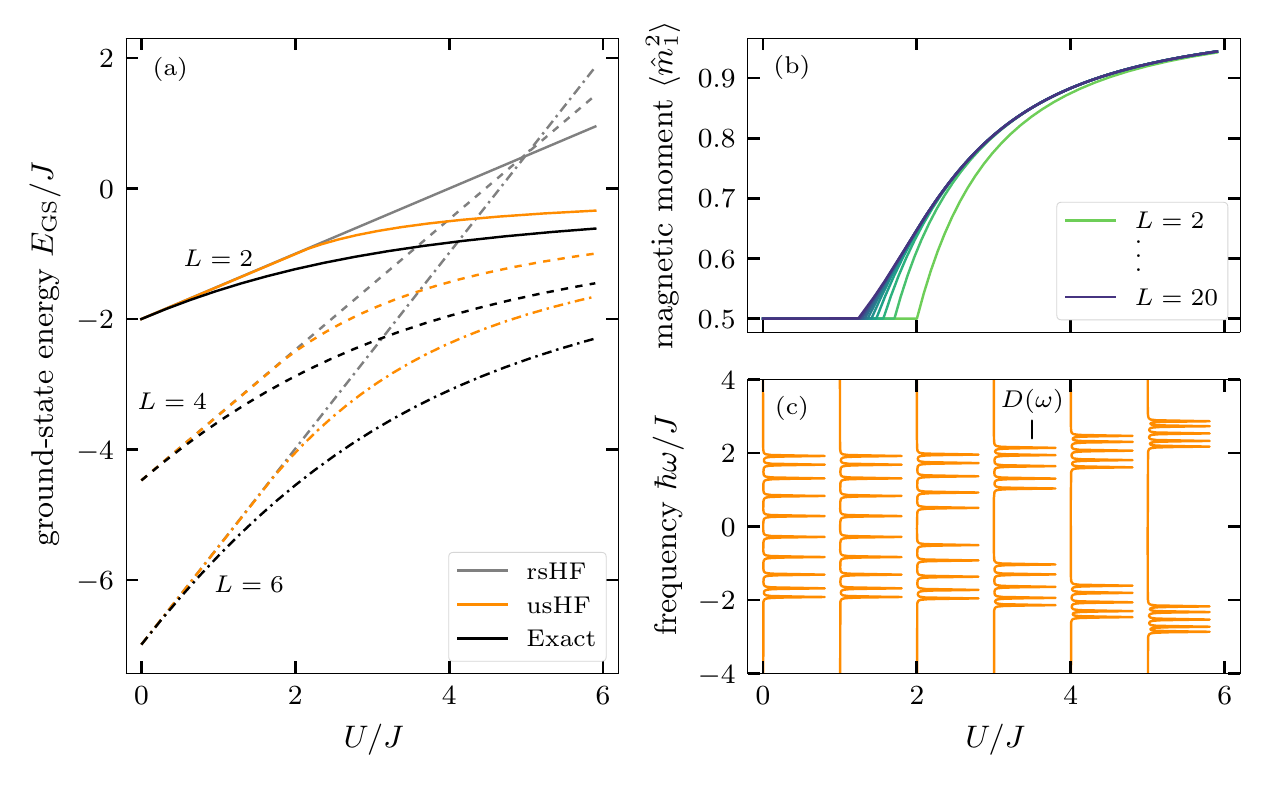}
\caption{(a) Comparison of the ground-state energy $E_\tn{GS}$ vs. interaction strength $U$ for rsHF, usHF, and the exact solution of the Hubbard Hamiltonian. Three finite Hubbard chains of length $L=2, 4, 6$ with open boundary conditions at half filling are considered. (b) Interaction dependence of the local magnetic moment $\langle \chat{m}^2 \rangle$, Eq.~(\ref{eq:moment}), on the first site of finite Hubbard chains of length $L=2,\ldots,20$ with open boundary conditions at half filling for usHF. (c) Interaction dependence of the DOS for a finite Hubbard chain of length $L=10$ with open boundary conditions at half filling for usHF. A Mott transition occures at $U\approx2J$.
}
\label{fig:hf}       %
\end{figure}

Since the exact ground state of the half-filled 1D Hubbard Hamiltonian is known to be spin-symmetric (i.e. paramagnetic) for systems with an even number of particles~\cite{lieb_two_1989,lieb_uniform_1993}, a logical prescription is to introduce spin symmetry also for approximate solutions like HF and SOA.
However, it is well known that, beyond a critical interaction strength $U_\textrm{c}$, unrestricted-spin HF (usHF) 
spontaneously
breaks spin symmetry resulting in an antiferromagnetic ground state. 
In the following we quantify the influence of this 
artificial
phase transition on important ground-state properties by comparing the performance of restricted-spin (rs) and unrestricted-spin (us) HF for finite one-dimensional Hubbard chains with open (hard-wall) boundary conditions. In Fig.~\ref{fig:hf}(a) the ground-state energy of three Hubbard clusters containing $L=2,4,6$ sites is plotted vs. the interaction strength $U$ for rsHF, usHF, and the result obtained by exact diagonalization of the Hamiltonian. The qualitative observations are similar for all three systems. In the limit of vanishing on-site interaction all three methods agree perfectly and show a linear increase of the ground-state energy with $U$. For interactions beyond $U\gtrsim1J$ the exact energy is reduced due to increasing correlations giving rise to mounting differences compared to the two HF solutions. In the case of rsHF, which by design fulfills the exact spin symmetry of the system, the linear increase of the ground-state energy is present for all values of $U$ resulting in a strong deviation from the exact result, for $U\gtrsim1J$. Most notably, since correlations are not included in rsHF, no Mott regime is observed in the presence of the on-site interaction. 

By contrast, removing the requirement of spin-symmetry (usHF) results in a lower ground-state energy for interactions beyond $U_\mathrm{c}\approx2J$ which approaches the exact value for $U\to\infty$. 
Additionally, the usHF density of states, shown in Fig.~\ref{fig:hf}(c) for a Hubbard chain of ten sites, is indicative that a correlation gap in the spectrum (i.e. a Mott transition) 
emerges for a critical interaction $U_\mathrm{c}$. However, since the usHF selfenergy accounts only for mean-field effects, the improved results for the Hubbard gap and the ground-state energy cannot be attributed to the effects of correlations. Instead, they are connected to the emergence of the antiferromagnetic state.
This becomes apparent when looking at the local magnetic moment on the outermost site of finite Hubbard chains depicted in Fig.~\ref{fig:hf}(b). The local magnetic moment is defined as 
\begin{align}
  \left\langle \chat{m}_i^2 \right\rangle = \left\langle \left( \chat{n}_{i,\uparrow} - \chat{n}_{i,\downarrow} \right)^2 \right\rangle = n_i - 2d_i\,,
\label{eq:moment}
\end{align}
with $n_i = n_{i,\uparrow} + n_{i,\downarrow}$ and the local double occupancy $d_i$ which contains a mean-field and a correlation part,
 \begin{align}
      d_i = n_{i,\uparrow}n_{i,\downarrow} + d^\mathrm{corr}_i\,.
 \end{align}
In the exact case where the spin densities are homogeneous, an increasing magnetic moment at high interaction strengths is caused by an increase of electronic correlations
leading to a negative $d^\mathrm{corr}_i$ and, thus, a decrease in the double occupancy.
However, for usHF, where $d^\mathrm{corr}_i \equiv 0$, the inhomogeneous spin-density distribution of the antiferromagnetic spin state mimicks the effect of additional correlations.\footnote{Note that the critical interaction $U_\mathrm{c}$ for which the symmetry-broken state emerges decreases with increasing length of the chain. An in-depth analysis on the exact value of $U_\mathrm{c}$ goes beyond the scope of this work.}
To summarize, removing the requirement of a homogeneous spin-symmetric ground state, as observed in the exact solution, allows HF simulations to achieve results for ground-state energies closer to the exact ones
and gives rise to a Hubbard gap of reasonable magnitude.

\section{Symmetries in Second-Born Approximation}
\label{s:soa}
\begin{figure}
 \includegraphics[width=\textwidth]{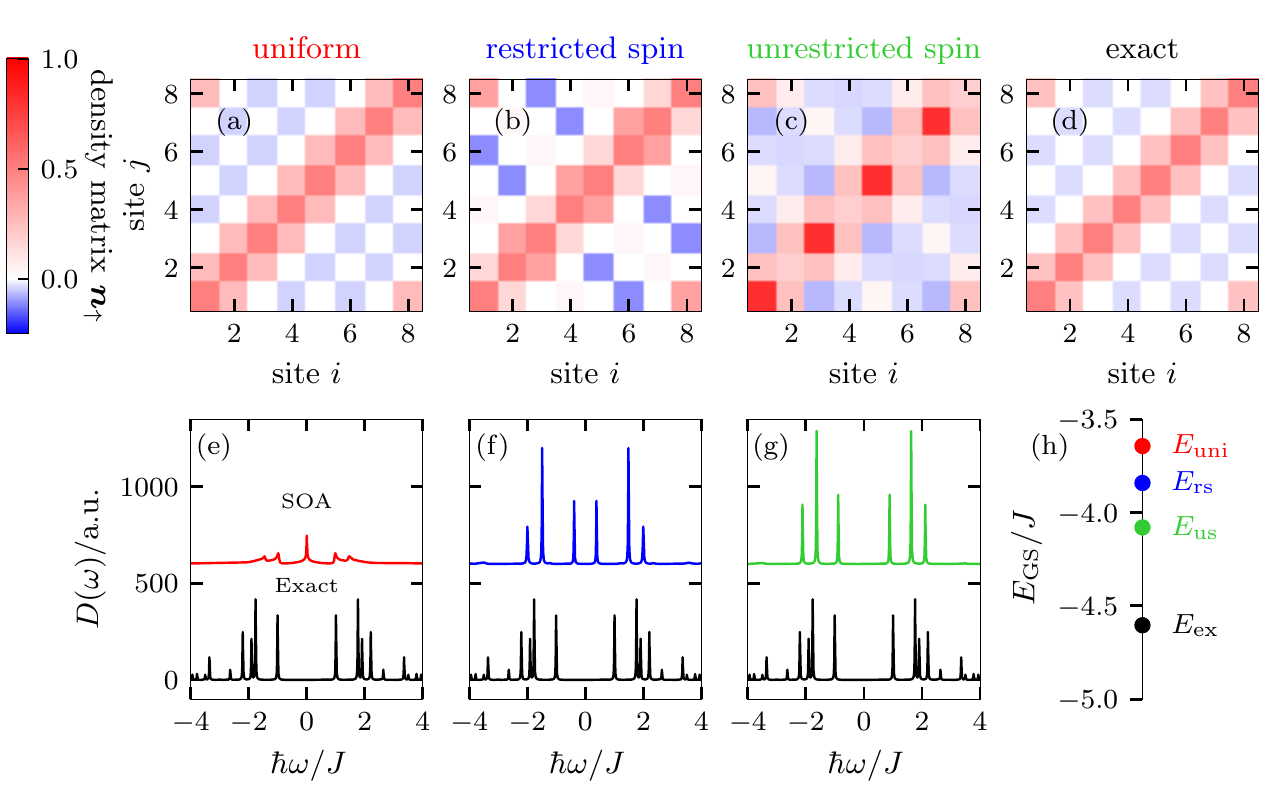}
\caption{Ground-state properties of a periodic, half-filled Hubbard chain of length $L=8$ and $U=4J$, within SOA. (a)-(d) Density matrix for a translationally invariant system (red), without imposing homogeneity but spin symmetry (blue), without both, homogeneity and spin restriction (green), and exact CI solution without restrictions (black).
(e)-(g) Spectral function (DOS) of the three approximations compared to CI results (``Exact''). (h) Total ground-state energy for the three cases, compared to the exact result.}
\label{fig:d-matrix}       %
\end{figure}

We next analyze the effect of imposing symmetry restrictions when including selfenergies with correlation effects (beyond HF). The first correction beyond HF that takes into account interparticle scattering is the second-order Born approximation (SOA). We thus consider, within SOA, the behavior of finite Hubbard chains with periodic boundary conditions for which the exact ground state is known to be both, spin symmetric and invariant under space translations~\cite{lieb_uniform_1993,lieb_two_1989,baeriswyl2013hubbard}. The effect of relaxing the aforementioned symmetry constraints %
is quantified by comparing results from the three SOA approaches I, II, and III, (i.e. for the uniform, restricted-spin and unrestricted-spin treatments) introduced in Sec.~\ref{s:gf_theory}; we will refer to these three treatments as uniSOA, rsSOA, and usSOA, respectively. For all three cases we compare in Fig.~\ref{fig:d-matrix} the spin-up density matrix (figure parts a-d), the DOS (e-g) and the ground-state energy (h) to the exact results for an 8-site Hubbard chain with 
periodic boundary conditions 
at $U=4J$. In the uniform case, the density matrix by design exhibits perfect translational symmetry
and the resulting checkerboard structure closely resembles the exact solution [cf. Fig.~\ref{fig:d-matrix} (a) and (d)].
At the same time, the ground-state energy of $-3.64J$ for uniSOA does not agree with the exact value of $-4.60J$. Similarly, the DOS shows poor qualitative agreement with the exact result, failing to reproduce the correct position of the peaks and, above all, the existence of a band gap.\\
Relaxing the requirement of translational symmetry (rsSOA) leads to unphysical inhomogeneities in the odd minor diagonals of the spin-up density matrix which is in contrast to the exact solution.\footnote{Note that, while the translational symmetry is broken, the spin symmetry is still fulfilled.} However, the ground-state energy is improved, to $-3.84J$, closer to the exact result. Additionally, in the DOS, a correlation-induced gap emerges at the Fermi energy in conjunction with an, in general, better qualitative agreement with the exact spectrum. Still, the rsSOA gap of $0.77J$ is less than half the size of the exact result of $2.01J$.\\
As a next step, we no longer enforce spin symmetry (usSOA) which results in an antiferromagnetic ground state indicated by the spin-density wave on the diagonal and the inhomogeneities on the even minor diagonals of the spin-up density matrix, shown in Fig.~\ref{fig:d-matrix}(c). This N\'eel state has an energy of $-4.08J$ which is in much better agreement with the exact value. In this case, the DOS nicely reproduces the position of the main peaks and the Hubbard gap of $1.77J$ is much closer to the exact value.\footnote{The remaining differences to the exact DOS, namely the missing high-energy satellites and the degenerate peak at $\sim \pm 1.8J$, can be attributed to the shortcoming of the SOA at the large interaction $U=4J$.}\\
Similar to the findings for the HF selfenergy presented in Sec.~\ref{s:hf}, removing the requirement of translation and spin symmetry for the SOA leads to a significant improvement for the ground-state energy and the DOS. This way the exact Mott gap can be remarkably well reproduced, even for relatively large interactions such as $U=4J$. 

\section{Multiple Solutions of the Dyson Equation in Second-Born Approximation}
\label{s:multi}

\begin{figure}
 \includegraphics{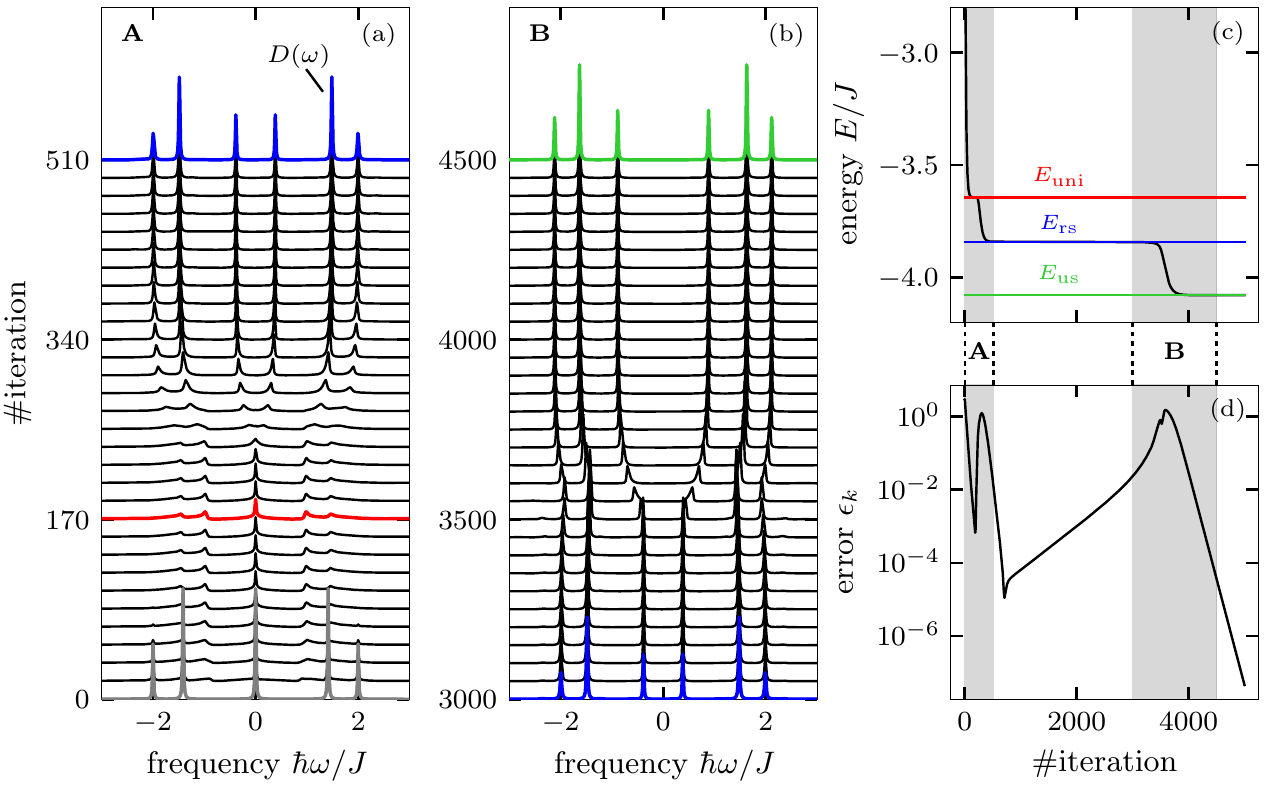}
\caption{Ground-state properties of a half-filled Hubbard chain of length $L=8$ and $U=4J$ for usSOA during the iteration procedure. The initial state is the rsHF solution of the system modified by a small deviation to initiate the symmetry breaking. (a)--(b) Evolution of the DOS during the iteration procedure, for two stages \textbf{A} and \textbf{B} of the iteration that are highlighted by the gray areas in (c) and (d). The colored spectra correspond to the respective states in Fig.~\ref{fig:d-matrix}. (c) Ground-state energy during the iteration procedure. The energies corresponding to the states of Fig.~\ref{fig:d-matrix} are shown with their respective color. (d) Relative error during the iteration procedure. 
}
\label{fig:iteration}       %
\end{figure}

Multiple solutions have been shown to be an inherent feature of self-consistent treatments of the 
Dyson equation \cite{tandetzky_multiplicity_2015,Stan_njp_2015, gunnarsson_breakdown_2017}. %
Usually, a self-consistency requirement is employed
in approximate treatments via e.g. perturbation theory. What we wish to discuss in this section 
is the connection between the insurgence of multiple approximate self-consistent solutions 
and the release of specific symmetry constraints.\\
To this end we solve Eq.~(\ref{eq:dyson}) with no symmetries enforced (usSOA), for a half-filled Hubbard chain of length $L=8$ with periodic boundary conditions and $U=4J$, i.e., for  the same system as in Sec.~\ref{s:soa}. The initial state of the iteration scheme is chosen to be the homogeneous, spin-restricted HF ground state where the density is modified by a small random perturbation of the order $10^{-5}$.\footnote{Recall that, without this small initial inhomogeneity, no broken spin symmetry would occur, even in spin-resolved calculations.} In Fig.~\ref{fig:iteration} the DOS (a-b), ground-state energy (c) and iteration error (d) are shown during the iterative procedure. The calculation starts from the slightly disturbed rsHF ground state with energy $-1.52J$ [not shown in Fig.~\ref{fig:iteration}(c)]. Within the first 170 iterations the system 
converges into the homogeneous state [cf. the DOS in Fig.~\ref{fig:d-matrix}(e)], as the relative iteration error drops to $10^{-3}$. However, at around 300 iterations the error increases and the system transitions into the inhomogeneous but spin-symmetric state [cf. the DOS in Fig.~\ref{fig:d-matrix}(f)]. This state %
appears to be even more stable, with the relative iteration error temporarily dropping to $10^{-5}$. However, after 3000 iterations, a final transition sets in, and the system arrives in the spin-asymmetric state [cf. the DOS in Fig.~\ref{fig:d-matrix}(g)]. The system remained in this state for the remainder of the iteration process, and the relative error of the calculation eventually reached the order of machine precision.\footnote{It should be noted that there is a little dip/kink in the relative error at around iteration 3500. This could hint to a possible forth viable state that we did not reach in our calculation.} 

While, during the iteration, the DOS evolves through the three states shown in Fig.~\ref{fig:d-matrix}, 
the total energy passes through the values of the respective states that are depicted in Fig.~\ref{fig:d-matrix}(h). This is shown in detail, as a function of the iteration number,  in Fig.~\ref{fig:iteration}.(c), cf. the colored lines. This shows that the three states, corresponding to the different symmetry restrictions discussed in Sec~\ref{s:soa}, can be reached by a single iterative solution of the Dyson equation when no symmetries are enforced. During the iteration, in the vicinity of each of the three states, the iteration error drops significantly (reaching a high degree of self-consistency). This suggests that all of them are solutions of the same Dyson equation, with only the unrestricted-spin state being absolutely numerically stable. A well conditioned iteration scheme will, ultimately, reach this minimum-energy state. Of course, the sequence of states reached on the way to the minimum depends on the choice of the initial state of the iteration and on details of the numerical procedure.

The present example is a direct illustration of Löwdin's symmetry dilemma discussed above and shows that a successive reduction of the symmetry of explored states may allow one to improve certain target properties of a system, such as the ground-state energy, also in a many-body calculation with the SOA selfenergy.
The behavior just discussed here is expected to be a specific facet of a general scenario underlying the search for multiple solutions of the Dyson equation that are a consequence of the nonlinear dependence of the collision integrals on the Green functions which is a general property of selfenergies beyond Hartree--Fock. These observations should give useful hints how to improve iterative solutions of the Dyson equation or similar nonlinear equations.

\section{Benchmarking Against DMRG}
\label{s:gap}

\begin{figure}
 \includegraphics[width=\textwidth]{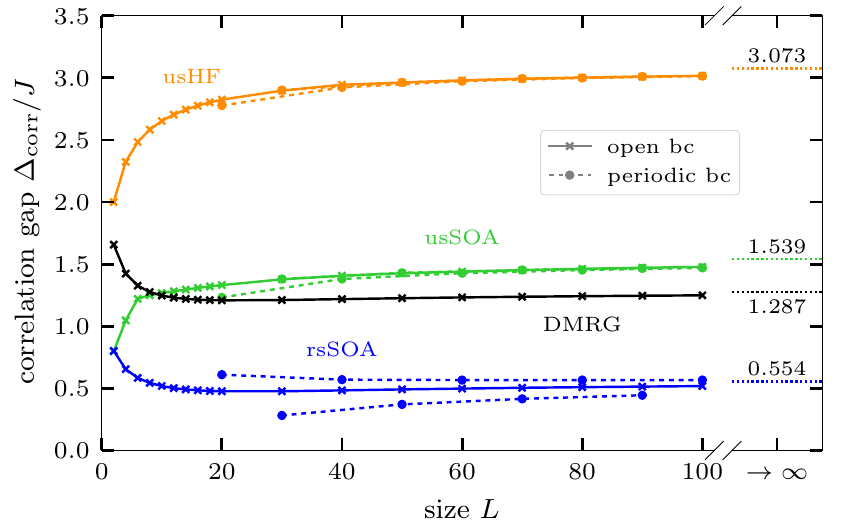}
\caption{Correlation band gap of a half-filled, one-dimensional Hubbard chain of finite length $L$ and $U=4J$ with open (full lines with crosses) and periodic (dashed lines with circles) boundary conditions. DMRG results are compared to different restricted spin (rs) and unrestricted spin (us) selfenergy approximations. The extrapolated results for the limit of the infinite Hubbard chain are shown as dotted lines on the right.}
\label{fig:gap}       %
\end{figure}

The existence of a Mott gap for large on-site interactions is one of the most important features of the Hubbard model. After analyzing the general properties of GFMBA simulations with HF and SOA selfenergies for the three methods (I)--(III), we now focus on their performance regarding the Hubbard gap, \refeqn{eq:gap},
in particular.
Since we are considering finite systems we are interested in the correlation part of the band gap which we define as
\begin{align}
    \Delta_\mathrm{corr} = \Delta - \Delta_\mathrm{rsHF}\,,
\end{align}
where $\Delta_\mathrm{rsHF}$ is the band gap obtained from a rsHF calculation that contains only the finite-size contribution. In the thermodynamic limit $\Delta_\mathrm{rsHF}$ vanishes, and $\Delta_\mathrm{corr} = \Delta$. In Fig.~\ref{fig:gap} we compare the correlation gap of finite Hubbard chains of varying length at $U=4J$ to the (exact) result obtained by DMRG (we employed the size-increasing scheme as in Refs.~\cite{PS1,PS2}). Additionally, we extrapolate the data to $L\to\infty$ where the DMRG result agrees with the Bethe-ansatz solution~\cite{essler_05}. In the case of the restricted-spin HF and the homogeneous SOA (uniSOA) state the Hubbard gap vanishes, cf. Fig.~\ref{fig:d-matrix}(d) for the latter.\\
In contrast, starting with open boundary conditions, rsSOA shows a finite gap and correctly predicts its qualitative dependence on the length of the system. However, since SOA captures only part of the correlation effects, the correct band gap is underestimated by $\sim0.7J$ for all system sizes. As discussed for the local magnetic moment in Sec.~\ref{s:hf} the antiferromagnetic ground state of the unrestricted spin methods can compensate shortcomings of the selfenergy approximations in treating correlations. In the case of usHF this results in the opening of a correlation gap on the mean-field level. However, for the large interaction of $U=4J$ the size of the gap is severely overestimated, especially, in the thermodynamic case, where usHF predicts a size of $3.073J$, as opposed to the exact value of $1.277J$. In contrast to the exact case, the correlated gap is monotonically increasing with the system length for the unrestricted-spin methods. Nevertheless, usSOA shows the best agreement with the exact gap out of all selfenergies and symmetry restrictions considered here. Especially for large system sizes, including the case of the infinite Hubbard chain, the deviation does not exceed $\sim0.25J$.\\
The results for periodic boundary conditions (shown as dashed lines in Fig.~\ref{fig:gap}) differ only slightly from the above observations. While, for small systems ($L<40$), there is a noticeable difference between both cases, the size of the correlated gap converges to the same value for larger systems. The speed of this convergence is considerably faster for the unrestricted-spin methods which possess an antiferromagnetic ground state.\\
Independent of the type of boundary conditions, giving up on the 
symmetries of the exact solution, 
the description of the Hubbard gap in finite systems, especially for the selfenergy in SOA, improves dramatically, even for large interaction strengths, %
again confirming Löwdin's symmetry dilemma. 

\section{Conclusion}
\label{s:conclusion}

Currently, there is great interest in the theoretical condensed-matter community
in devising approaches for strongly correlated systems,
i.e for those systems where a description based on an independent-particle picture is 
qualitatively inadequate. 

In this work, we have looked at a certain aspect of the problem, namely the interplay of
symmetry constraints and electronic correlations. 
Specifically, using finite Hubbard chains at half filling as a case study for strongly correlated systems,
we investigated the role of symmetry requirements in many-body perturbation Green 
function theory (GFMBA), where approximations
of increasing complexity can be systematically devised for the many-body  
selfenergy.

In the literature, GFMBA is often seen as an ill-suited conceptual paradigm to describe 
strong correlations, because of, e.g., possible multiple self-consistent approximate solutions, 
or uncertainty about the convergence radius of the perturbation expansion.
While not calling into question this point of view, in this study we have used GFMBA to address
the interplay of symmetry and correlation effects. To our knowledge this is a point that,
irrespective of the methodology used, has received little systematic attention so far. 
For our GFMBA description of Hubbard systems, we used the second-Born approximation 
which accounts for electronic correlations at lowest perturbative order in a ``skeleton-diagram'' sense. 
However, for the sake of comparison we also presented results from an Hartree--Fock treatment.

Already at the HF level, our comparison of spin-symmetry-restricted and unrestricted 
self-consistent solutions indicates that one is facing a so-called {\it L\"owdin symmetry dilemma}
for the one-dimensional Hubbard model: namely, 
the violation of spin-symmetry can lead to a spectral function and ground-state energy that are 
closer to the exact ones than those obtained when spin symmetry is imposed. 

Our results suggest that this behavior is robust 
against the inclusion of electronic correlations within GFMBA. For our self-consistent SOA treatment
of finite periodic Hubbard chains at half-filling, in addition to spin symmetry we also
considered translation symmetry.
While the exact solution corresponds to densities which are both spin-projection symmetric and
translationally invariant, the violation of these properties in SOA leads to results for the ground-state energy
and the local spectral function which are remarkably close to the exact ones.
This is particularly so when using the unrestricted-spin symmetry SOA: in this case 
the exact value of the Hubbard energy gap is surprisingly well produced, within $\sim 15\%$, also
for strong on-site interaction $U=4J$.
Finally, an interesting overall trait of our results is that the symmetry-restricted states 
are in fact solutions (albeit metastable) to the unrestricted Dyson equation: starting from a specifically 
crafted initial state, the self-consistency iteration dynamics 
passes through the symmetry-constrained states before reaching the unrestricted ground state.

Thus, altogether our work illustrates a direct connection between symmetry constraints and solution multiplicity in GFMBA, adding to the already available body of knowledge on the behaviour of multiple solutions for the Dyson equation.  As possible future directions, an obvious point to address is if more complex selfenergy approximations, such as the  $T$ matrix and $GW$ approximations, would give improved results in 
symmetry-lifted treatments as well. Other straightforward extensions would be the exploration of the case of  
higher dimensions (where, however, an increased complexity is expected due to a 
richer structure of the phase diagram) and electron occupancies different from half filling. 

More in general  (and very much in the spirit of L\"owdin's original lines of thought), it could be of some
interest to see if it is possible (and what happens) when concretely profiting from the symmetry dilemma
within GFMBA, by restoring symmetry at the end via projection techniques. This procedure, in different
variants and extensions, has been already used in the literature for wavefunctions, typically 
starting from HF symmetry-unrestricted states, e.g., Ref.~\cite{Yannouleas_2007}. Pursuing the same strategy
within GFMBA would allow one to explore  the feasibility
of  
convenient ways  to access some important physical quantities 
in strongly correlated systems.

\printbibliography

\end{document}